\PassOptionsToPackage{unicode}{hyperref}
\PassOptionsToPackage{hyphens}{url}
\PassOptionsToPackage{dvipsnames,svgnames,x11names}{xcolor}
\documentclass[
  12pt,
]{interact}
\usepackage{xcolor}
\usepackage{amsmath,amssymb}
\usepackage{iftex}
\ifPDFTeX
  \usepackage[T1]{fontenc}
  \usepackage[utf8]{inputenc}
  \usepackage{textcomp} 
\else 
  \usepackage{unicode-math} 
  \defaultfontfeatures{Scale=MatchLowercase}
  \defaultfontfeatures[\rmfamily]{Ligatures=TeX,Scale=1}
\fi
\usepackage{lmodern}
\ifPDFTeX\else
\fi
\IfFileExists{upquote.sty}{\usepackage{upquote}}{}
\IfFileExists{microtype.sty}{
  \usepackage[]{microtype}
  \UseMicrotypeSet[protrusion]{basicmath} 
}{}
\usepackage{setspace}
\makeatletter
\@ifundefined{KOMAClassName}{
  \IfFileExists{parskip.sty}{%
    \usepackage{parskip}
  }{
    \setlength{\parindent}{0pt}
    \setlength{\parskip}{6pt plus 2pt minus 1pt}}
}{
  \KOMAoptions{parskip=half}}
\makeatother
\makeatletter
\ifx\paragraph\undefined\else
  \let\oldparagraph\paragraph
  \renewcommand{\paragraph}{
    \@ifstar
      \xxxParagraphStar
      \xxxParagraphNoStar
  }
  \newcommand{\xxxParagraphStar}[1]{\oldparagraph*{#1}\mbox{}}
  \newcommand{\xxxParagraphNoStar}[1]{\oldparagraph{#1}\mbox{}}
\fi
\ifx\subparagraph\undefined\else
  \let\oldsubparagraph\subparagraph
  \renewcommand{\subparagraph}{
    \@ifstar
      \xxxSubParagraphStar
      \xxxSubParagraphNoStar
  }
  \newcommand{\xxxSubParagraphStar}[1]{\oldsubparagraph*{#1}\mbox{}}
  \newcommand{\xxxSubParagraphNoStar}[1]{\oldsubparagraph{#1}\mbox{}}
\fi
\makeatother

\usepackage{longtable,booktabs,array}
\usepackage{calc} 
\usepackage{etoolbox}
\makeatletter
\patchcmd\longtable{\par}{\if@noskipsec\mbox{}\fi\par}{}{}
\makeatother
\IfFileExists{footnotehyper.sty}{\usepackage{footnotehyper}}{\usepackage{footnote}}
\makesavenoteenv{longtable}
\usepackage{graphicx}
\makeatletter
\newsavebox\pandoc@box
\newcommand*\pandocbounded[1]{
  \sbox\pandoc@box{#1}%
  \Gscale@div\@tempa{\textheight}{\dimexpr\ht\pandoc@box+\dp\pandoc@box\relax}%
  \Gscale@div\@tempb{\linewidth}{\wd\pandoc@box}%
  \ifdim\@tempb\p@<\@tempa\p@\let\@tempa\@tempb\fi
  \ifdim\@tempa\p@<\p@\scalebox{\@tempa}{\usebox\pandoc@box}%
  \else\usebox{\pandoc@box}%
  \fi%
}
\def\fps@figure{htbp}
\makeatother

\providecommand{\tightlist}{%
  \setlength{\itemsep}{0pt}\setlength{\parskip}{0pt}}

\usepackage[]{natbib}
\usepackage{booktabs}
\usepackage{longtable}
\usepackage{array}
\usepackage{multirow}
\usepackage{wrapfig}
\usepackage{float}
\usepackage{colortbl}
\usepackage{pdflscape}
\usepackage{tabu}
\usepackage{threeparttable}
\usepackage{threeparttablex}
\usepackage[normalem]{ulem}
\usepackage{makecell}
\usepackage{xcolor}
\usepackage{orcidlink}
\makeatletter
\@ifpackageloaded{caption}{}{\usepackage{caption}}
\AtBeginDocument{%
\ifdefined\contentsname
  \renewcommand*\contentsname{Table of contents}
\else
  \newcommand\contentsname{Table of contents}
\fi
\ifdefined\listfigurename
  \renewcommand*\listfigurename{List of Figures}
\else
  \newcommand\listfigurename{List of Figures}
\fi
\ifdefined\listtablename
  \renewcommand*\listtablename{List of Tables}
\else
  \newcommand\listtablename{List of Tables}
\fi
\ifdefined\figurename
  \renewcommand*\figurename{Figure}
\else
  \newcommand\figurename{Figure}
\fi
\ifdefined\tablename
  \renewcommand*\tablename{Table}
\else
  \newcommand\tablename{Table}
\fi
}
\@ifpackageloaded{float}{}{\usepackage{float}}
\floatstyle{ruled}
\@ifundefined{c@chapter}{\newfloat{codelisting}{h}{lop}}{\newfloat{codelisting}{h}{lop}[chapter]}
\floatname{codelisting}{Listing}

\makeatother
\makeatletter
\@ifpackageloaded{caption}{}{\usepackage{caption}}
\@ifpackageloaded{subcaption}{}{\usepackage{subcaption}}
\makeatother
\usepackage{bookmark}
\IfFileExists{xurl.sty}{\usepackage{xurl}}{} 
\hypersetup{
  pdftitle={Inside Out: Externalizing Assumptions in Data Analysis as Validation Checks},
  pdfauthor={H. Sherry Zhang; Roger D. Peng},
  pdfkeywords={diagnostic, logic regression, data analysis assumptions},
  colorlinks=true,
  linkcolor={blue},
  filecolor={Maroon},
  citecolor={Blue},
  urlcolor={Blue},
  pdfcreator={LaTeX via pandoc}}

\title{Inside Out: Externalizing Assumptions in Data Analysis as
Validation Checks}
\author{H. Sherry Zhang$\textsuperscript{1}$, Roger D.
Peng$\textsuperscript{1}$}

\thanks{CONTACT: H. Sherry
Zhang. Email: \href{mailto:huize.zhang@austin.utexas.edu}{\nolinkurl{huize.zhang@austin.utexas.edu}}. Roger
D.
Peng. Email: \href{mailto:roger.peng@austin.utexas.edu}{\nolinkurl{roger.peng@austin.utexas.edu}}. }
\begin{document}
\captionsetup{labelsep=space}
\maketitle
\textsuperscript{1} Department of Statistics and Data
Sciences, University of Texas at Austin, Texas, United States
\begin{abstract}
In data analysis, unexpected results often prompt researchers to revisit
their procedures to identify potential issues. While some researchers
may struggle to identify the root causes, experienced researchers can
often quickly diagnose problems by checking a few key assumptions. These
checked assumptions, or expectations, are typically informal, difficult
to trace, and rarely discussed in publications. In this paper, we
introduce the term \emph{analysis validation checks} to formalize and
externalize these informal assumptions. We then introduce a procedure to
identify a subset of checks that best predict the occurrence of
unexpected outcomes, based on simulations of the original data. The
checks are evaluated in terms of accuracy, determined by binary
classification metrics, and independence, which measures the shared
information among checks. We demonstrate this approach with a toy
example using step count data and a generalized linear model example
examining the effect of particulate matter air pollution on daily
mortality.
\end{abstract}
\begin{keywords}
\def\sep{;\ }
diagnostic, logic regression, data analysis assumptions\sep 
diagnostic, logic regression, data analysis assumptions
\end{keywords}

\setstretch{1.15}
\section{Introduction}\label{introduction}

In data analysis, analysts often rely on prior knowledge or domain
expertise to form expectations and assess whether results align with
them. When results deviate from these expectations, experienced analysts
check assumptions made about the data during the analysis, but often do
so without discussing the underlying reasoning in the publications.
While publication of analysis code and data is now a common requirement
for the sake of reproducibility \citep{peng2011reproducible}, the
published code alone is often insufficient for understanding the thought
process behind an analysis. Code corresponding to published analyses
often reflects the final decisions made about analysis and does not
reveal the decisions or assumptions about the data during the analysis.
Thus, a reader looking at published code can often be left with many
questions about why certain choices were made.

We might gain insight into analysts' thought processes by speaking with
them directly or watching them work via screencast videos they produce,
such as TidyTuesday screencast videos or think-aloud type studies
\citep[e.g.][]{gu2024data}. However, direct observation of analysis is
not scalable and may not always be feasible; creating educational
screencast videos requires significant effort from the analysts.
Ideally, there could be a way to explicitly communicate these thought
processes in the data analysis to others. Even better, if these
expectations were machine-readable, we could analyze them and learn from
the analysis itself. For example, we could answer questions about
whether the checks also apply to other researchers analyzing new data in
the same context, whether they reflect common practices in the field, or
whether they are specific to the data or analysis at hand. Externalizing
these thought processes can improve the trustworthiness of the analyses
\citep{yu2024veridical}.

The trustworthiness of an analysis depends not only on the robustness of
the methods used but also on a deep understanding of the data. This
corresponds to the third step identified in
\citet{broderick_toward_2023}, where trust can break down if the
assumptions underlying an algorithm are not satisfied in practice. In
the past, data were often collected directly by analysts through
carefully designed experiments, however, contemporary analyses often
rely on data curated by government or research institutions, where some
data characteristics may be unknown to analysts. When unexpected
outcomes arise, they are often later traced back to implicit assumptions
made by the analysts - assumptions that were unknowingly violated by the
idiosyncrasies in the data. By externalizing these assumptions through
analysis validation checks, we aim to make it easier to identify and
diagnose such unexpected results.

In this paper, we conceptualize these implicit expectations as
\emph{analysis validation checks}, which allows us to examine the
assumptions made about the data during an analysis. We then introduce a
procedure to identify the subset of checks that best predict the
occurrence of unexpected outcomes. The procedure, based on simulations
of the original data, compares the accuracy and independence of
different combinations of analysis validation checks. Accuracy is
determined using binary classification metrics (precision and recall)
from a logic regression model \citep{ruczinski_logic_2003}, while
independence measures the shared information among checks. The proposed
workflow offers a numerical guarantee that the analysis will produce the
expected results, assuming the assumptions about the data generating
mechanism hold.

The rest of the paper is organized as follows:
Section~\ref{sec-lit-review} reviews the concepts of diagnosing
unexpected outcomes and general data quality checks.
Section~\ref{sec-plan} introduces the concept of analysis validation
checks, illustrated with a toy example based on step count data.
Section~\ref{sec-method} describes the procedure that selects the best
subset of checks for understanding the occurrence of unexpected
outcomes. Section~\ref{sec-pm10-mortality} applies this procedure to a
larger example that estimates the effect of particulate matter air
pollution on daily mortality. Section~\ref{sec-discussion} summarizes
the paper and discusses a few key considerations.

\section{Related Work}\label{sec-lit-review}

\subsection{Diagnosing unexpected outcomes in data
analysis}\label{diagnosing-unexpected-outcomes-in-data-analysis}

The concept of framing data analysis as a sense-making process was
originally presented by \citet{grolemund_cognitive_2014} based on
seminal work by \citet{wild1999statistical}. Key to any sense-making
process is a model for the world (i.e.~expectations for what we might
observe) and observed data with which we can compare our expectations.
If there is a significant deviation between what we observe and our
expectations, then a data analysis must determine what is causing that
deviation. A naive approach would be to update our model for the world
to match the data, under the assumption that the initial expectation was
incorrect. However, experienced analysts know that the reality can be
more nuanced than that, with errors occurring in data collection or data
processing that can have an impact on final results.

The skill of diagnosing unexpected data analysis results is not one that
has received significant attention in the statistics literature. While
the concept of diagnosis is often embedded in model checking or data
visualization techniques, systematic approaches to identifying the root
cause of a general unexpected analysis result are typically not
presented \citep{peng2022perspective}. Furthermore, if interesting
information is discovered through model checking or data visualization,
there is no formal way to document such discoveries for the next analyst
to consider. \citet{peng_diagnosing_2021} proposed a series of exercises
for training students in data analysis to diagnose different kinds of
analysis problems such as coding errors or outliers. They provide a
systematic approach involving working backwards from the analysis result
to identify potential causes. There are parallels here to the concept of
debugging and testing in software engineering
\citep{donoghue2021teaching}. For example, \citet{li2019towards} found
that experienced engineers were generally able to identify problems in
code faster than novices, and that the ability to debug code required
knowledge that cut across different domains.

If it is true that the speed with which data analysts can identify
problems with an analysis is related to their experience working with a
given type of data, then there is perhaps room to improve the analytic
process by externalizing the aspects that an analyst learns through
experience. That way, inexperienced analysts could examine the thought
process of an experienced analyst and learn to identify factors that can
cause unexpected results to occur.

\subsection{Data analysis checks}\label{data-analysis-checks}

The concept of data quality has been studied in the literature, with
earlier work focusing on defining frameworks -- such as dimensions,
attributes, and measures -- to improve data quality in database and
information systems
\citep{wang1996beyond, batini2009methodologies, 6204995, woodall2014classification, cai2015challenges, 8642813}.
More recently, data validation has been incorporated into frameworks
like Google TensorFlow \citep{polyzotis2019data} to ensure the quality
of data for training machine learning models, as well as for supporting
business decision-making \citep{schelter2018automating}. With the
increasing availability of open data in scientific research, the users
of the data are often not the original data collectors and may not be
aware of all the details or nuances of the data. This encourages
researchers to conduct data quality checks before beginning the analytic
process, helping to avoid unexpected discoveries later on. In R,
packages like \texttt{skimr} \citep{skimr} and \texttt{dataMaid}
\citep{dataMaid} provide basic data screening and data quality reports,
while packages such as \texttt{assertr} \citep{assertr},
\texttt{validate} \citep{validate}, and \texttt{pointblank}
\citep{pointblank} focus on providing infrastructures that allow users
to define customized data quality checks.

The literature on data quality typically focuses on the intrinsic or
inherent quality of the data themselves, rather than the data's
relationship to any specific data analysis. So for example, if a column
in a data table is expecting numerical data, but we observe a character
value in one of the entries, then that occurrence would trigger some
sort of data quality check. This type of quality check can be triggered
without any knowledge of what the data will ultimately be used for.
However, for a given analysis, we may require specific aspects of the
data to be true because they affect the result being computed.
Conversely, certain types of poor quality data may have little impact on
the ultimate result of an analysis (e.g.~data that are missing
completely at random). Defining data quality in terms of what may affect
a specific analysis outcome or result has the potential to open new
avenues for defining data checks and for building algorithms for
optimizing the collection of checks defined for a specific analysis.

\section{Analysis validation checks}\label{sec-plan}

Analysis validation checks are assumptions made by the analysts during
the analysis, framed as explicit checks that return a binary TRUE or
FALSE value based on the data. Inspired by the concept of data
validation checks \citep{validate}, which are designed to ensure that
datasets meet expected quality before the analysis begins, analysis
validation checks reverse the approach: they validate the assumptions
about the data necessary for the analysis to produce the \emph{expected
results}, as defined by the analyst. The focus on expected results
allows analysis validation checks to encompass a wide range of checks,
such as data quality, variable distributions and outliers, bivariate and
multivariate relationships, and contextual information relevant to the
analysis.

Our proposed analysis validation checks provide insights into an
analyst's thought process and offer the following benefits:

\begin{enumerate}
\def\labelenumi{\arabic{enumi}.}
\tightlist
\item
  Serve as clear checkpoints to support the replication or application
  of methods to (new) data by programmatically communicating the
  requirements or assumptions made of the data;
\item
  Align assumptions among researchers from different domain backgrounds
  who may have different expectations about the data;
\item
  Improve analysis transparency, reproducibility, and trustworthiness by
  externalizing a key part of the analysis process; and
\item
  Quantify the effectiveness of analysis checks for predicting the
  expected outcome (see Section~\ref{sec-method});
\end{enumerate}

In addition to the above benefits, the development and publication of
analysis checks have the potential to help students, inexperienced
analysts, and junior researchers develop the skills needed to diagnose
unexpected analysis results for a given type of data because the
assumptions made about the data are made transparent. The analysis
checks can serve as a basis for new analysts to have conversations about
the data they are analyzing and to develop a better understanding of the
potential data generation process.

\subsection{A Toy Example}\label{sec-toy}

Consider a 30-day step count experiment in a public health setting.
Subjects are instructed to walk at least 8,000 steps each day, with an
expected average of 9,000 steps, tracked by a step counter app. With
data of this nature, we may expect there to be occasional ``low'' days
due to factors such as unfavorable weather conditions limiting outdoor
activities. We may also expect ``high'' days after outdoor activities or
intense workouts. Given the requirements of the study, we form our
expectation as follows:

\begin{quote}
Expectation: The average step count of a given subject is between
\([8,500, 9,500]\)
\end{quote}

To diagnose potential reasons why this expectation might fail, we can
establish a few analysis validation checks in anticipation of seeing the
data. {[}TODO: Add guidelines for designing the checks{]} For example,
we can check the quantile of the step count, if more than a third of the
days fall below 8,000, or more than a third exceed 10,000 steps, this
could indicate an excess of low-count or high-count days. Similarly, we
may expect the standard deviation of the step count not to be overly
large. These considerations yield the following three analysis
validation checks that \emph{fail} when:

\begin{itemize}
\tightlist
\item
  Check 1: the 60\% quantile of the observed step counts is greater than
  10,000
\item
  Check 2: the 40\% quantile of the observed step counts is less than
  8,000, and
\item
  Check 3: the standard deviation of the observed step counts exceeds
  2,500.
\end{itemize}

The cutoff values for these checks would presumably be chosen based on
prior experience with these kinds of data, but could also be optimized
using the method presented in the next section.

\begin{figure}

\centering{

\pandocbounded{\includegraphics[keepaspectratio]{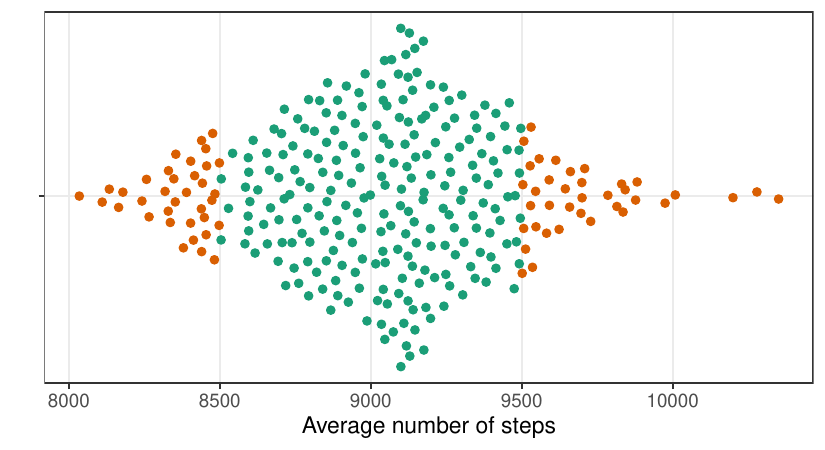}}

}

\caption{\label{fig-step-count}Beeswarm plot of average step counts
across 300 simulated 30-day periods. Each point represents the average
step count from one simulation. Similar to a boxplot or violin plot, the
beeswarm plot also displays the distribution of each individual data
point. The orange points indicate instances where the average step count
falls outside the {[}8,500, 9,500{]} interval, representing an
unexpected outcome in this scenario.}

\end{figure}

To simulate this data, three normal distributions are used for the daily
step counts: \(\mathcal{N}(6,000, 200)\) for low days,
\(\mathcal{N}(12,000, 200)\) for high days, and
\(\mathcal{N}(9,000, 300)\) for typical days. The number of low and high
days can be simulated from a Poisson distribution with \(\lambda = 8\).
Figure~\ref{fig-step-count} displays average step count across 300
simulated 30-day periods.

\section{Method}\label{sec-method}

While not all analysis validation checks are equally important, some may
be more useful at diagnosing unexpected outcomes than others. This
section introduces a procedure to identify a subset of analysis
validation checks that best predict the occurrence of unexpected
outcomes.

\begin{figure}

\centering{

\includegraphics[width=1\linewidth,height=\textheight,keepaspectratio]{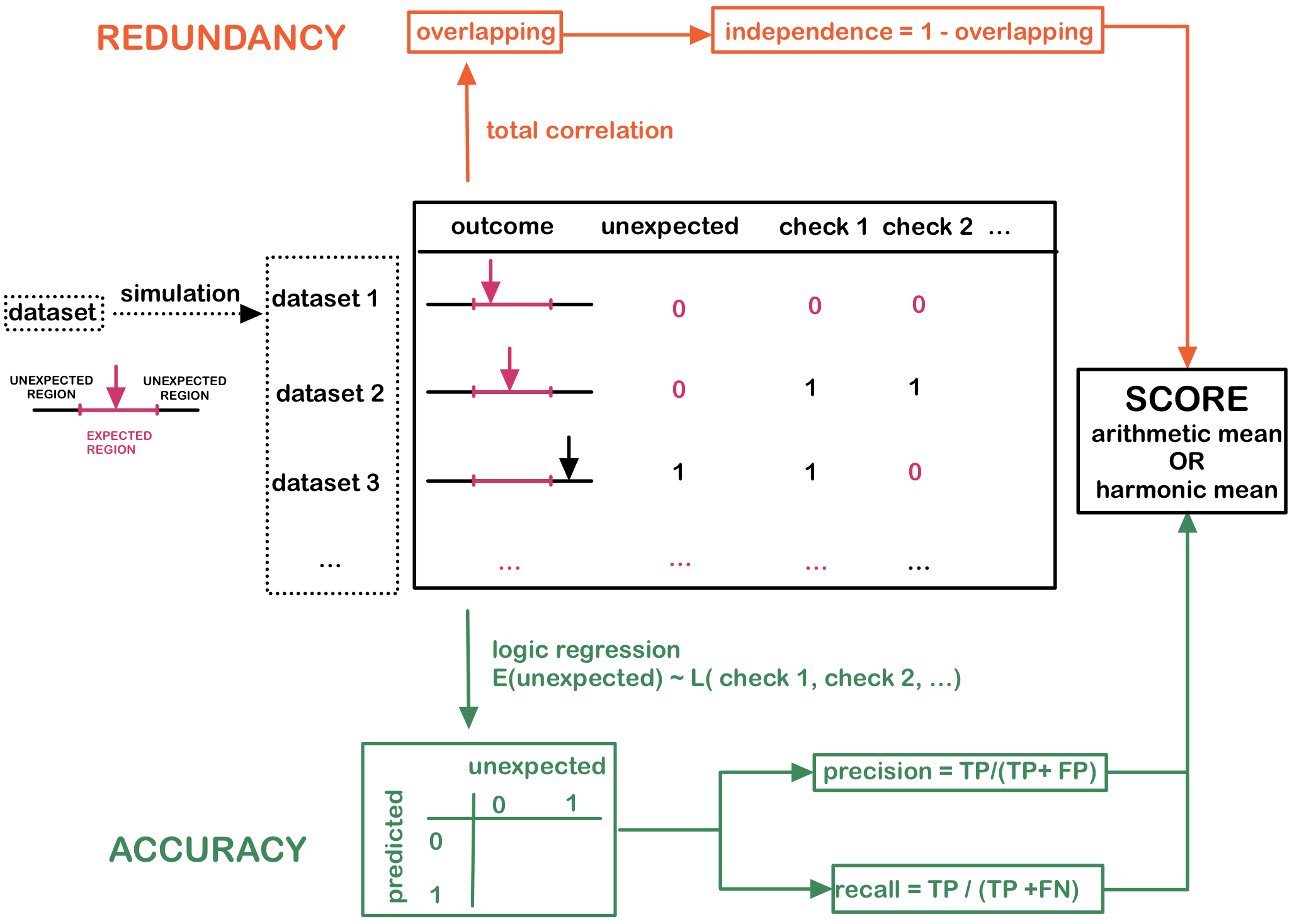}

}

\caption{\label{fig-metric-calc}Overview of the procedure to quantify
the effectiveness of analysis validation checks. The procedure involves
simulating replicates of the data, applying the analysis, and running
the analysis validation checks on each replicate to determine whether
the outcome meets the expectation (the unexpected column) and whether
each check passes (check 1, check 2, \ldots). The simulation results are
then passed to two branches: the accuracy branch calculates the
precision and recall of the checks from a logic regression prediction,
while the redundancy branch calculates the mutual information and
independence of the checks. These three metrics are combined to produce
a single metric that quantifies the effectiveness of the checks in
diagnosing unexpected outcomes in the analysis.}

\end{figure}

The approach generates simulated datasets from the observed data to
compute the outcome and the analysis validation checks (coded as 1 for
unexpected and 0 as expected). This then allows us to apply a logic
regression tree to link the outcome and the check results to construct a
Boolean expression to best predict the unexpected outcomes.
Figure~\ref{fig-metric-calc} provides an overview of the process.

The specific strategy to simulate the data depends on the problem and
the nature of the data at hand. Simulated data can be generated using
prior domain knowledge, summary statistics derived from the observed
data, or statistical techniques such as training-testing splits or
bootstrap sampling. While each simulated data may not be independent of
the original data - as is the case with bootstrap samples - it should
adequately capture the range of possible idiosyncrasies present in the
data. Failure to do so can reduce the ability of the resulting Boolean
expression to predict unexpected outcomes.

\subsection{Accuracy}\label{accuracy}

From the simulated data, the accuracy branch in
Figure~\ref{fig-metric-calc} refers to the set of checks' ability to
detect unexpected outcomes accurately. To relate multiple checks to the
outcome, a logic regression model \citep{ruczinski_logic_2003} is used.
Originally developed for SNP micro-array data, logic regression predicts
classification or regression outcomes by constructing Boolean
combinations, \(\mathcal{L}(X_1, X_2, \cdots, X_k)\), where \(X_i\) are
binary predictors. The logical operations allowed to construct
\(\mathcal{L(\cdot)}\) include AND (\(\land\)), OR (\(\lor\)), and NOT
(\(\neg\)) and an example of the Boolean combination of
\(\mathcal{L(\cdot)}\) is \(X_1 \text{ AND } (X_2 \text{ OR } X_3)\). In
our binary-binary classification problem, the logic regression model can
be written as follows:

\[E(Y) = \mathcal{L}(X_1, X_2, \cdots, X_k),\]

where the outcome, \(Y\), indicates whether the outcome is
\emph{unexpected} (1) or \emph{expected} (0), and the analysis
validation checks, \(X_1, X_2, \cdots, X_k\), are labeled as 1 if they
fail and 0 if they pass.

This objective function is optimized using a simulated annealing
algorithm to minimize the misclassification error and the following six
moves are permitted during the optimization to grow or prune the tree
(as detailed in Figure 2 of \citet{ruczinski_logic_2003}): 1) replacing
a leaf node, 2) replacing an operator, 3) growing a branch, 4) pruning a
branch, 5) splitting a leaf node, and 6) deleting a leaf node.
Predictions from the logic regression model are then used, along with
the observed outcome, to calculate the precision and recall of the
optimized Boolean combination of checks. Compared to other tree-based
methods for binary-binary prediction, the Boolean combinations from the
logic regression model produce a tree structure that can be directly
interpreted as the possible combination of checks leading to an
unexpected outcome, without the need to invert the tree as required in
classic tree-based recursive partitioning methods.

\subsection{Independence}\label{independence}

While checks may score high on predictive accuracy, they may be
correlated and less useful to provide actions to act on the analysis.
This could happen if a set of checks are all tangentially related to the
cause of the unexpected results, but none addresses the root cause. To
quantify the redundancy among the checks, we use total correlation,
\(C\), a multivariate generalization of mutual information that captures
the amount of shared information across multiple variables. We define
the normalized total correlation, \(C'\), as

\[C'(X_1, X_2, \cdots, X_k) = \frac{C(X_1, X_2, \cdots, X_k)}{\sum_{i= 1}^k H(X_i)} = \frac{\sum_{i= 1}^k H(X_i) - H(X_1, X_2, \cdots, X_k)}{\sum_{i= 1}^k H(X_i)}, \]
where \(H(X_i)\) is the entropy of the \(i\)-th check and
\(H(X_1, X_2, \cdots, X_k)\) is the joint entropy of checks
\(\{X_1, X_2, \cdots, X_k\}\). The amount of independent information,
\(\eta\), can be defined as:

\[\eta(X_1, X_2, \cdots, X_k) = 1 - C'(X_1, X_2, \cdots, X_k) = \frac{H(X_1, X_2, \cdots, X_k)}{\sum_{i= 1}^k H(X_i)}.\]

To scale this quantity to the range \([0,1]\) for consistency with other
metrics (i.e., precision and recall), we define the independence metric
as:

\[\text{independence} = \frac{\eta - 1/k}{1 - 1/k}= \frac{H(X_1, X_2, \cdots, X_k)/\sum_{i= 1}^k H(X_i) - 1/k}{1 - 1/k}.\]

An independence metric value of 1 indicates that all checks are
independent: \(H(X_1, X_2, \cdots, X_k)=\sum_{i= 1}^k H(X_i)\) and each
check provides unique information to diagnose the unexpected outcome
(the trivial case is when there is only one check). A value of 0
indicates the checks are all identical:
\(H(X_1, X_2, \cdots, X_k)- \sum_{i= 1}^k H(X_i) = H(X_i)/ k H(X_i) = 1/k\).

The three metrics (precision, recall, and independence) can be combined
into a single metric using the arithmetic mean, harmonic mean, or
quadratic mean. The differences among these means are minimal when the
three metrics are similar. However, as the differences among the metrics
increases, the harmonic mean tends to produce the smallest overall
score, as it penalizes low values, while the quadratic mean tends to
produce the largest score by rewarding higher values more. For simple
interpretation of the score, the arithmetic mean is preferred, while in
applications where the difference between precision, recall, and
independence need to be penalized or rewarded more, the harmonic and
quadratic mean should be considered.

\subsection{Toy Example Revisited}\label{toy-example-revisited}

\begin{figure}

\centering{

\pandocbounded{\includegraphics[keepaspectratio]{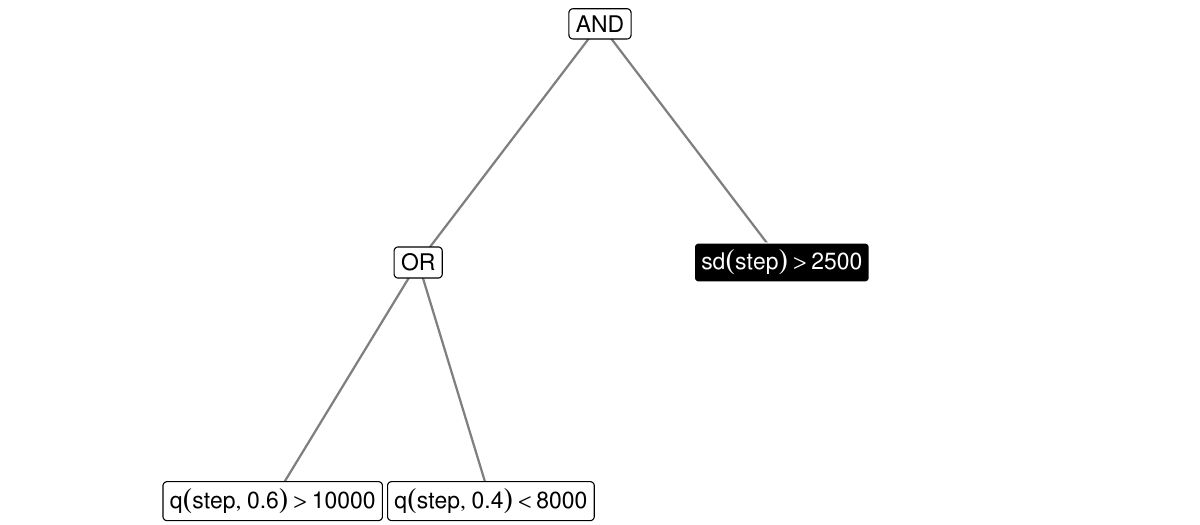}}

}

\caption{\label{fig-logic-reg}Logic regression model fitted to the three
checks. The model suggests the rule (quantile(step, 0.6) \textgreater{}
10,000 OR quantile(step, 0.4) \textless{} 8,000) AND (NOT sd(step)
\textgreater{} 2,500). The NOT operator applied to
\texttt{sd(step)\ \textgreater{}\ 2,500} is colored with a black
background to distinguish it from other checks.}

\end{figure}

Returning to the step count example introduced in Section~\ref{sec-toy},
the logic regression model is fitted to the three validation checks
described previously to predict the outcome, whether the average number
of steps fall within the {[}8,500, 9,500{]} interval.
Figure~\ref{fig-logic-reg} shows the best-fitting logic regression model
as

\begin{quote}
(quantile(step, 0.6) \textgreater{} 10,000 OR quantile(step, 0.4)
\textless{} 8,000) AND (NOT sd(step) \textgreater{} 2,500)
\end{quote}

We would predict an unexpected outcome in the analysis if the standard
deviation of the step count is not too large (2,500) and either the 60\%
quantile of the step count exceeds 10,000 or the 40\% quantile of the
step count falls below 8,000.

Table~\ref{tbl-logic-reg} presents the calculated precision, recall, and
independence for the three individual checks and the check rule found by
the logic regression. The harmonic and arithmetic means are included to
combine the three measures. The results show that the three checks
produced by the logic regression can accurately predict 86.1\% cases of
all \emph{predicted unexpected results} in the simulation data.
Furthermore, 43.1\% of all \emph{actual unexpected results} were in fact
observed to be unexpected.

\begin{table}

\caption{\label{tbl-logic-reg}Precision, recall, and independence
calculated for each individual check and the logic regression check
rule. The harmonic and arithmetic means of the three metrics are
included to evaluate the quality of the checks in diagnosing unexpected
step counts (more than five days with fewer than 8,000 steps).}

\centering{

\centering
\resizebox{\ifdim\width>\linewidth\linewidth\else\width\fi}{!}{
\begin{tabular}{>{\raggedright\arraybackslash}p{19em}rrrrr}
\toprule
Checks & Precision & Recall & Independence & Harmonic & Arithmetic\\
\midrule
Check 1: q(step, 0.6) > 10000 & 0.319 & 0.575 & 1.000 & 0.511 & 0.631\\
Check 2: q(step, 0.4) < 8000 & 0.264 & 0.613 & 1.000 & 0.467 & 0.626\\
Check 3: sd(step) > 2500 & 0.153 & 0.289 & 1.000 & 0.273 & 0.481\\
Logic regression: (check 1 OR check 2) AND (not check 3) & 0.861 & 0.431 & 0.818 & 0.637 & 0.703\\
Comparison: (check 1) AND (not check 3) & 0.278 & 0.870 & 0.881 & 0.510 & 0.676\\
\addlinespace
Regression tree & 0.542 & 0.780 & 0.818 & 0.689 & 0.713\\
\bottomrule
\end{tabular}}

}

\end{table}%

For comparison, we include a simplified rule, (quantile(step, 0.6)
\textgreater{} 10,000) AND (NOT sd(step) \textgreater{} 2,500). Although
the checks on the 60th quantile and standard deviation are less
correlated -- reflected in a higher independence score -- this pair is
less accurate at predicting the unexpected outcome, resulting in lower
harmonic and arithmetic mean scores. The regression tree produces a
similar prediction to the logic regression, however, we argue that the
logic regression tree shown in Figure~\ref{fig-logic-reg} is more
interpretable for our purposes because it provides a direct
representation of which combinations of analysis checks lead to
unexpected outcomes. The logic regression tree is also directly
comparable to other diagnostic techniques, which we discuss further in
Section~\ref{sec-discussion}.

\section{Application}\label{sec-pm10-mortality}

In the study of the health effects of outdoor air pollution, one area of
interest is the association between short-term, day-to-day changes in
particulate matter air pollution and daily mortality counts. Substantial
work has been done to study this question and to date, there appears to
be strong evidence of an association between particulate matter less
than 10 \(\mu\)g/m\(^3\) in aerodynamic diameter (PM10) and daily
mortality from all non-accidental causes \citep{samet2000fine}. In the
following example, we use the problem of studying PM10 and mortality
along with data from the National Morbidity, Mortality, and Air
Pollution Study (NMMAPS) to demonstrate how our analysis validation
checks described in Section~\ref{sec-method} can be applied. In addition
to providing a more substantial problem for our methods, this example
also demonstrates how the procedure presented in
Section~\ref{sec-method} can be used to select cutoff values in the
analysis checks to diagnose an unexpected PM10 coefficient from the
generalized linear model. The dataset that we use to develop our
simulations and analysis checks is from New York City, and contains
daily PM10, all-cause (non-accidental) mortality, and average
temperature values from 1992--2000. We then will apply the analysis
checks to similar data from 39 other cities from the NMMAPS study.

The typical approach to studying the association between PM10 and
mortality is to apply a generalized linear model (Poisson family with a
log link) to relate daily mortality counts to daily measures of PM10.
Based on previous work and the range of effect sizes published in the
literature, an analyst might expect the coefficient for PM10 in this GLM
to lie between \([0, 0.005]\), after adjusting for daily temperature
\citep{samet2000fine, welty2005acute}. Note that the relatively simple
modeling approach being used here is primarily for demonstration;
typically far more sophisticated semi-parametric approaches are used in
the literature \citep{peng2006model}.

Multiple factors can affect the estimated PM10 coefficient, such as the
strength of the correlation between mortality and PM10, or between
mortality and temperature. Outliers in the variables can also leverage
the coefficient. While these are possible factors that could affect the
analysis result, it is not clear what the cutoff values for these checks
should be to determine a failure. Here we consider a list of checks in
Table~\ref{tbl-checks} with varied cutoff values:

\begin{longtable}[]{@{}l@{}}
\caption{List of checks considered for the generalized linear model of
mortality on PM10 and temperature. The checks are based on the sample
size, correlation between mortality and PM10, correlation between
mortality and temperature, and univariate outlier detection. Multiple
cutoff values are specified for each check to determine a
failure.}\label{tbl-checks}\tabularnewline
\toprule\noalign{}
The check fails (encoded as 1) if \ldots{} \\
\midrule\noalign{}
\endfirsthead
\toprule\noalign{}
The check fails (encoded as 1) if \ldots{} \\
\midrule\noalign{}
\endhead
\bottomrule\noalign{}
\endlastfoot
Mortality-PM10 correlation less than \(-0.05\) \\
Mortality-PM10 correlation less than \(-0.03\) \\
Mortality-PM10 correlation greater than \(0.03\) \\
Mortality-PM10 correlation greater than \(0.05\) \\
Mortality-temperature correlation greater than \(-0.3\) \\
Mortality-temperature correlation greater than \(-0.35\) \\
Mortality-temperature correlation greater than \(-0.4\) \\
Mortality-temperature correlation greater than \(-0.45\) \\
Outliers are present in the variable PM10 \\
Outliers are present in the variable mortality \\
\end{longtable}

\subsection{Data Simulation}\label{data-simulation}

To generate replicates of the dataset, we first generate the correlation
matrix of the three variables (PM10, mortality, and temperature) in a
grid and then use a Gaussian copula to generate a multivariate normal
distribution based on the specified correlation matrix and sample size.
The multivariate normal distribution is transformed using the normal CDF
before the inverse CDF of the assumed distributions of the three
variables is applied. To determine the appropriate distribution of each
variable, various distributions are fitted and compared. This includes
Poisson and negative binomial for mortality; gamma, log-normal,
exponential, Weibull, and normal for PM10 and temperature; and beta for
PM10 after rescaling the data to \([0,1]\).

\begin{figure}

\centering{

\pandocbounded{\includegraphics[keepaspectratio]{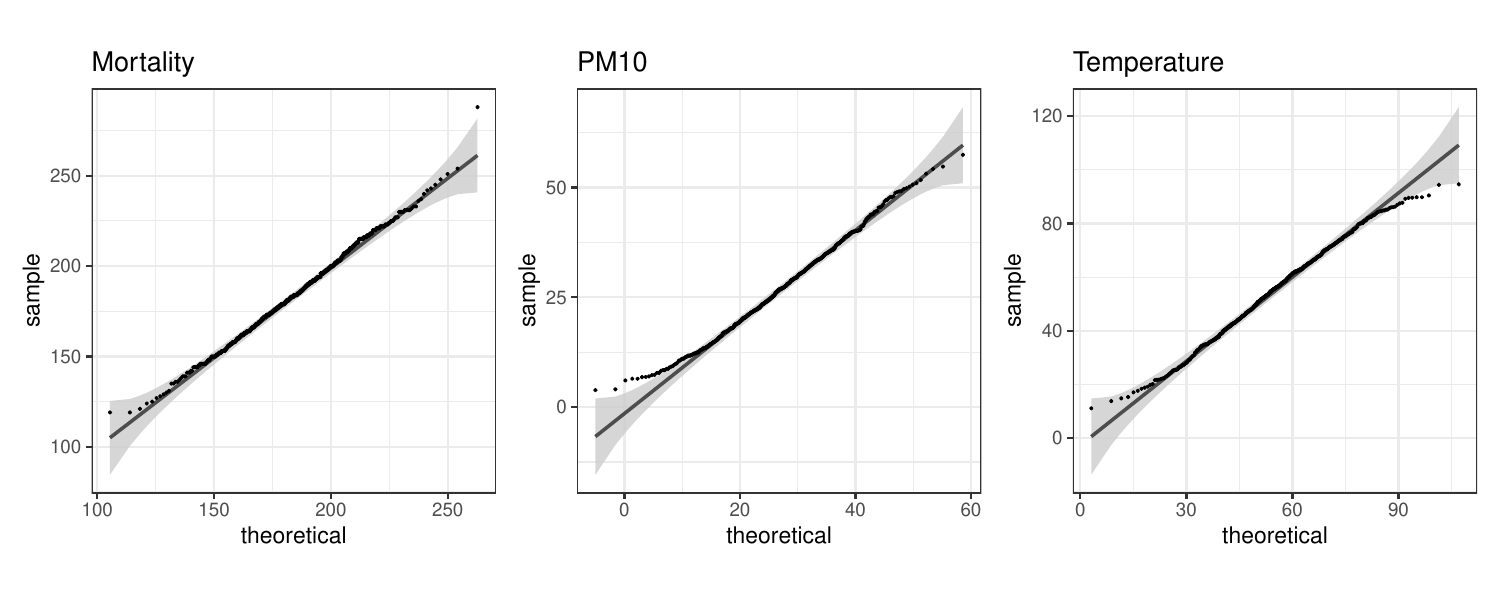}}

}

\caption{\label{fig-dist-fit}QQ-plot of the goodness-of-fit of selected
distributions used to generate simulated data for mortality, PM10, and
temperature. Left: Mortality as a negative binomial distribution (size =
74, mean = 183), Middle: PM10 is modeled using a rescaled beta
distribution with shape parameters 4.21 and 11.67, multiplied by 100 to
match the observed range. Right: Temperature is modeled using a Weibull
distribution (shape = 3.8, scale = 61)}

\end{figure}

To ensure reasonable similarity to observed data, we use characteristics
of the observed New York City dataset to refine our simulations. AIC is
used to determine the best distribution fit for each variable with the
QQ-plot presented in Figure~\ref{fig-dist-fit} to evaluate the fit. AIC
suggests a negative binomial distribution for mortality, a beta
distribution for PM10 (multiplied by 100 to recover the original scale),
and a Weibull distribution for temperature. To include the potential
effect of outliers, we add a single outlier to the data for both the
mortality and PM10 variables.

\begin{figure}

\centering{

\pandocbounded{\includegraphics[keepaspectratio]{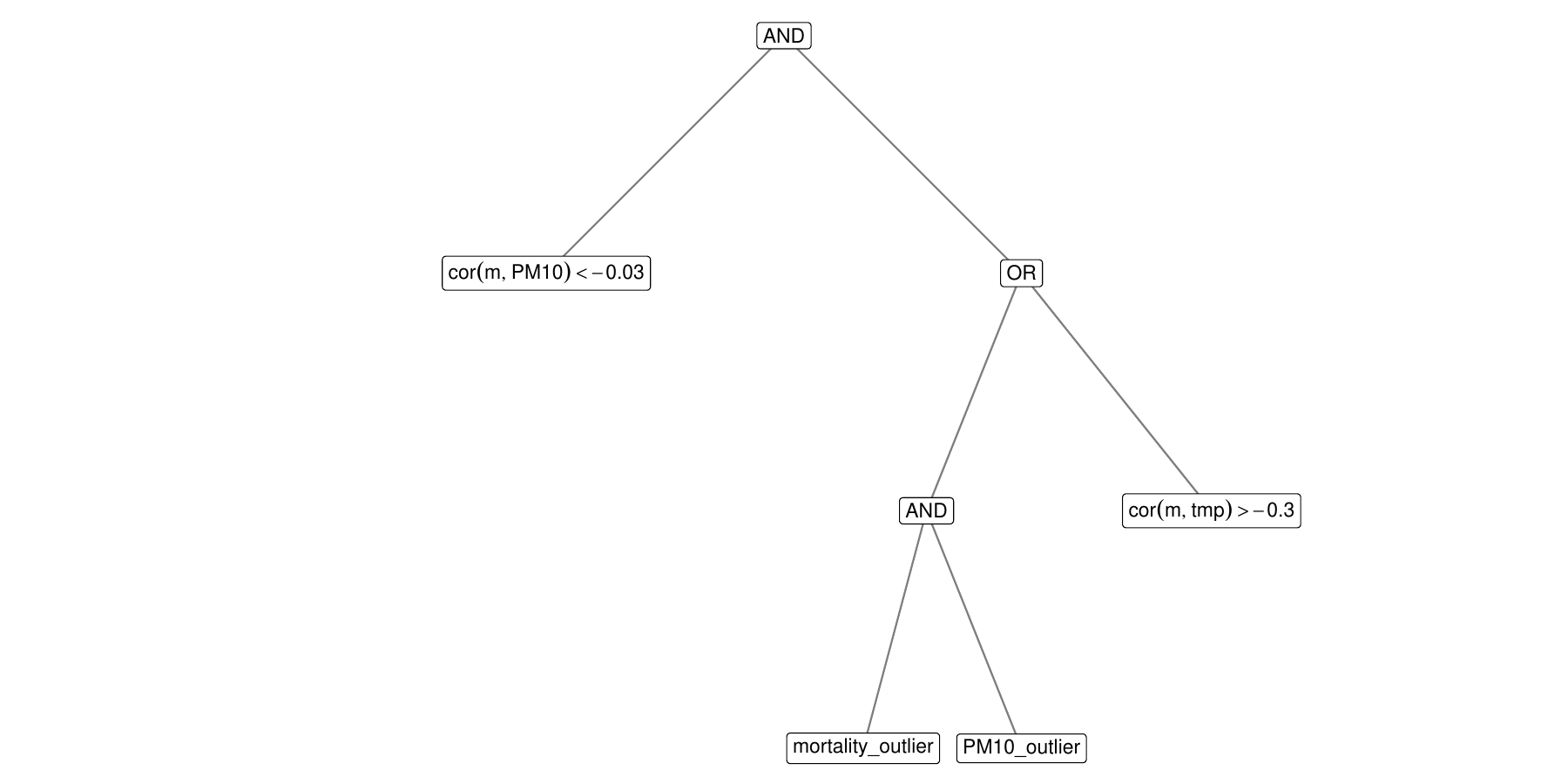}}

}

\caption{\label{fig-linear-reg-tree}Logic regression model fitted to the
fourteen checks and the outcome expectation (unexpected) as the response
variable. The model suggests the relationship: (mortality-PM10
correlation \textless{} \(-0.03\)) AND (mortality-temperature
correlation \textgreater{} \(-0.3\) OR (there exists mortality outlier
AND there exists PM10 outlier)) to predict the unexpected PM10
coefficient.}

\end{figure}

A logic regression is fitted using all variations of the checks in
Table~\ref{tbl-checks} to predict whether the PM10 coefficient is
unexpected. Given the outcome imbalance (expected vs.~unexpected),
inverse weights proportional to the number of observations in each
outcome are applied during the logic regression fit.
Figure~\ref{fig-linear-reg-tree} shows the optimal logic regression tree
from the fitted model. Precision, recall, and independence score, along
with their harmonic and arithmetic means are presented in
Table~\ref{tbl-linear-reg}.

\begin{table}

\caption{\label{tbl-linear-reg}Precision, recall, and independence
metrics derived from the logic regression model and each individual
check, along with harmonic and arithmetic means.}

\centering{

\centering
\resizebox{\ifdim\width>\linewidth\linewidth\else\width\fi}{!}{
\begin{tabular}{>{\raggedright\arraybackslash}p{18em}rrrrr}
\toprule
Checks & Precision & Recall & Independence & Harmonic & Arithmetic\\
\midrule
Check 1: cor(m, PM10) < -- 0.03 & 0.551 & 0.894 & 1.000 & 0.763 & 0.815\\
Check 2: cor(m, tmp) > -- 0.35 & 0.281 & 0.556 & 1.000 & 0.472 & 0.612\\
Check 3: mortality outlier & 0.302 & 0.775 & 1.000 & 0.536 & 0.692\\
Check 4: PM10 outlier & 0.305 & 0.766 & 1.000 & 0.537 & 0.690\\
Logic regression: (check 1) AND ((check 2) OR (check 3 AND check 4)) & 0.876 & 0.722 & 0.812 & 0.798 & 0.803\\
\bottomrule
\end{tabular}}

}

\end{table}%

As indicated in Figure~\ref{fig-linear-reg-tree}, the logic regression
model picks up the following cutoff values for each type of check:

\begin{itemize}
\tightlist
\item
  mortality-PM10 correlation less than \(-0.03\)
\item
  mortality-temperature correlation greater than \(-0.3\)
\item
  PM10 data contains outliers detected by the univariate outlier
  detection
\item
  mortality data contains outliers detected by the univariate outlier
  detection
\end{itemize}

The fitted logic regression model is

\begin{quote}
(cor(mortality, PM10) \textless{} -- 0.03) AND (cor(mortality, tmp)
\textgreater{} -- 0.3 OR (mortality outlier AND PM10 outlier))
\end{quote}

This generates a precision of 0.876 and a recall of 0.722 for predicting
the unexpected PM10 coefficient. As shown in
Figure~\ref{fig-linear-reg-tree}, there is no single analysis check in
the tree that predicts an unexpected outcome. Rather, at least three
checks in the tree must be TRUE in order for the model to predict an
unexpected outcome. Given the high independence of the checks
(Table~\ref{tbl-linear-reg}), this suggests that unexpected results are
only likely after multiple anomalies are observed in the data.

\subsection{Additional Cities}\label{additional-cities}

In order to test our development of analysis validation checks based on
the New York City data, we incorporate data from 39 additional cities
from the original NMMAPS study and apply our checks to those data. These
cities (along with New York City) represent approximately the largest 40
metropolitan areas in the United States by population. The complete list
of cities used are presented in Table~\ref{tbl-all-cities}. For each
city, we had daily data on all-cause mortality, PM10, and average
temperature for the years 1992--2000.

\begin{table}

\caption{\label{tbl-all-cities}Results from full analysis of 40 cities
from the NMMAPS air pollution and mortality study. In the table, a ``T''
indicates that a check failed and an ``F'' indicates that a check
passed.}

\centering{

\centering
\resizebox{\ifdim\width>\linewidth\linewidth\else\width\fi}{!}{
\begin{tabular}{lrccrrcc}
\toprule
City & \makecell[l]{PM10\\Estimated\\Coefficient} & \makecell[l]{Outcome\\Observed\\Unexpected} & \makecell[l]{Logic Regression\\Predicted\\Unexpected} & \makecell[l]{Correlation:\\Mortality-\\Temperature} & \makecell[l]{Correlation:\\Mortality-\\PM10} & \makecell[l]{Mortality\\Outlier} & \makecell[l]{PM10\\Outlier}\\
\midrule
Atlanta & --0.0008 & T & T & --0.26 (T) & --0.16 (T) & F & F\\
Austin & --0.0030 & T & T & --0.12 (T) & --0.11 (T) & F & T\\
Baltimore & 0.0018 & F & F & --0.21 (T) & 0.01 (F) & T & T\\
Boston & 0.0019 & F & F & --0.16 (T) & 0.02 (F) & F & T\\
Buffalo & 0.0028 & F & F & --0.23 (T) & --0.02 (F) & F & T\\
\addlinespace
Chicago & 0.0011 & F & F & --0.38 (F) & --0.02 (F) & T & T\\
Cincinnati & 0.0011 & F & T & --0.27 (T) & --0.06 (T) & T & T\\
Cleveland & 0.0010 & F & F & --0.30 (F) & --0.02 (F) & T & T\\
Columbus & 0.0004 & F & T & --0.20 (T) & --0.07 (T) & F & T\\
Denver & 0.0010 & F & F & --0.23 (T) & 0.06 (F) & F & T\\
\addlinespace
Detroit & 0.0011 & F & F & --0.30 (T) & 0.04 (F) & F & T\\
Dallas/Fort Worth & 0.0016 & F & T & --0.30 (F) & --0.03 (T) & T & T\\
Honolulu & 0.0006 & F & F & --0.17 (T) & 0.05 (F) & T & T\\
Houston & 0.0007 & F & F & --0.22 (T) & --0.03 (F) & F & T\\
Indianapolis & 0.0014 & F & F & --0.13 (T) & 0.02 (F) & F & T\\
\addlinespace
Kansas City & 0.0022 & F & F & --0.20 (T) & 0.02 (F) & F & T\\
Los Angeles & 0.0011 & F & F & --0.47 (F) & --0.01 (F) & T & T\\
Las Vegas & 0.0008 & F & F & --0.27 (T) & 0.05 (F) & T & T\\
Memphis & --0.0007 & T & T & --0.16 (T) & --0.10 (T) & F & T\\
Miami & --0.0004 & T & F & --0.16 (T) & --0.02 (F) & F & T\\
\addlinespace
Milwaukee & 0.0026 & F & F & --0.22 (T) & 0.09 (F) & F & T\\
Minneapolis/St. Paul & 0.0009 & F & F & --0.30 (F) & --0.01 (F) & T & T\\
New York & 0.0022 & F & F & --0.46 (F) & --0.01 (F) & T & T\\
Oakland & 0.0008 & F & F & --0.28 (T) & 0.04 (F) & F & T\\
Orlando & --0.0008 & T & T & --0.21 (T) & --0.05 (T) & F & T\\
\addlinespace
Philadelphia & 0.0021 & F & F & --0.27 (T) & 0.11 (F) & T & T\\
Phoenix & 0.0009 & F & F & --0.44 (F) & 0.07 (F) & T & T\\
Pittsburgh & 0.0009 & F & T & --0.36 (F) & --0.08 (T) & T & T\\
Riverside & 0.0001 & F & T & --0.27 (T) & --0.11 (T) & T & T\\
Sacramento & 0.0005 & F & F & --0.30 (T) & 0.05 (F) & T & T\\
\addlinespace
Salt Lake City & --0.0004 & T & T & --0.19 (T) & --0.03 (T) & F & T\\
San Antonio & --0.0021 & T & T & --0.22 (T) & --0.16 (T) & F & T\\
San Bernardino & 0.0006 & F & T & --0.28 (T) & --0.09 (T) & T & T\\
San Diego & 0.0011 & F & F & --0.33 (F) & 0.03 (F) & T & T\\
San Jose & 0.0006 & F & F & --0.23 (T) & 0.08 (F) & F & T\\
\addlinespace
Seattle & 0.0003 & F & F & --0.24 (T) & 0.06 (F) & F & T\\
Santa Ana/Anaheim & 0.0005 & F & T & --0.26 (T) & --0.04 (T) & T & T\\
St. Petersburg & 0.0012 & F & F & --0.31 (F) & 0.01 (F) & F & T\\
Tampa & --0.0003 & T & T & --0.20 (T) & --0.04 (T) & F & T\\
Tucson & 0.0028 & F & F & --0.31 (F) & 0.18 (F) & T & T\\
\bottomrule
\end{tabular}}

}

\end{table}%

We applied each of the analysis validation checks and the fitted logic
regression model separately to each of the cities' data to make
predictions of whether the outcome will be unexpected or not. We also
applied the generalized linear model to the data from each of the cities
to estimate the association between PM10 and mortality in each city,
adjusted for temperature. The estimates of the association and the
results of each of the analysis validation checks for each city are
shown in Table~\ref{tbl-all-cities}.

\begin{table}

\caption{\label{tbl-accuracy}Summary of the observed and the predicted
unexpected PM10 coefficient results from the 40 NMMAPS cities using the
logic regression model.}

\centering{

\begin{tabular}{lrr}
\toprule
\multicolumn{1}{c}{} & \multicolumn{2}{c}{Predicted} \\
\cmidrule(l{3pt}r{3pt}){2-3}
Observed & FALSE & TRUE\\
\midrule
FALSE & 25 & 7\\
TRUE & 1 & 7\\
\bottomrule
\end{tabular}

}

\end{table}%

It is clear from Table~\ref{tbl-all-cities} that some of the estimated
PM10 coefficients are outside the expected range {[}0, 0.005{]}. Eight
of the 40 cities had negative estimated coefficients and were therefore
considered unexpected by our original criterion. (It is perhaps worth
noting that none of the unexpected outcomes was in the positive
direction.) Table~\ref{tbl-accuracy} summarizes the prediction accuracy
of the logic regression model for the 40 cities. The model correctly
predicts the PM10 coefficient status of 32 cities, producing an accuracy
rate of 80\%. Among the 8 cities with an unexpected outcome, the model
correctly identifies 7 (Atlanta, Austin, Memphis, Orlando, Salt Lake
City, San Antonio, Tampa), giving a recall of 88\% (only Miami were not
properly classified by the logic regression model). These cities were
flagged due to failures in both the mortality-PM10 correlation and
mortality-temperature correlation analysis checks. Out of the 14 cities
with positive predictions from the model, 7 cities (Cincinnati,
Columbus, Dallas/Fort Worth, Pittsburgh, Riverside, San Bernardino,
Santa Ana/Anaheim) were false positives, resulting in a precision of
only 50\%. This precision value is substantially lower than what was
estimated by the simulation procedure (see Table~\ref{tbl-linear-reg})
and suggests that the simulation does not adequately capture some
features of the data generation process.

\section{Discussion}\label{sec-discussion}

In this paper, we have developed an approach to using analysis
validation checks to externalize the assumptions about the data and
analysis tools made during the data analysis process. These checks can
serve as a useful summary of the analyst's thought process and can
describe how characteristics of the data may lead to unexpected
outcomes. Using logic regression, we can develop a graphical summary of
the analysis validation checks as well as use the logic regression
fitting process to choose the optimal set of checks. The logic
regression model can also be used to develop summaries of the precision
and recall of the collection of analysis validation checks in predicting
the likelihood of an unexpected outcome. We demonstrated our method on
an example relating daily mortality to outdoor air pollution data. The
results from that example could be used to inform future analyses of air
pollution and health data, perhaps in other cities or locations.

In Section~\ref{sec-pm10-mortality} we used the analysis checks in a
diagnostic manner to examine the eight cities whose PM10 coefficients
were unexpected. There, we found that seven of the cities had a
mortality-PM10 correlation that was more negative than expected while
also having a mortality-temperature correlation that was somewhat larger
(less negative) than expected. What this result implies for the broader
analysis depends on a number of factors, but the interplay between
mortality, PM10, and temperature may warrant further investigation
\citep[see e.g.][]{welty2005acute}. Note that simply because the PM10
coefficient estimates were unexpected by our criterion, we do not mean
to imply that they are ``wrong'' in any sense. Indeed, negative
estimated coefficients have been found in other studies of this nature
\citep{bell2004ozone}. Rather, in this type of analysis, it may be that
the interval for expected results needs to be revised. Further, it may
be necessary to incorporate other outputs from the analysis, such as
measures of uncertainty.

An interesting connection can be drawn between our logic regression
trees and a tool used in systems engineering known as a fault tree. A
fault tree is used for conducting a structured risk assessment and has a
long history in aviation, aerospace, and nuclear power applications
\citep{vesely1981fault}. A fault tree is a graphical tool that describes
the possible combinations of causes and effects that lead to an anomaly.
At the top of the tree is a description of an anomaly. The subsequent
branches of the tree below the top event indicate possible causes of the
event immediately above it in the tree. The tree can then be built
recursively until we reach a root cause that cannot be further
investigated. Each level of the tree is connected together using logic
gates such as AND and OR gates. The leaf nodes of the tree indicate the
root causes that may lead to an anomaly. While the logic regression
trees are not identical to fault trees, they share many properties, such
as the tree-based structure and the indicator of root causes at the leaf
nodes. Perhaps more critically, both serve as graphical summaries of the
assumptions made in a problem and the specific violations of those
assumptions that could lead to an unexpected result. While fault trees
are often used to discover the root cause of an anomaly after it occurs,
an important use case for fault trees is to develop a comprehensive
understanding of a system \emph{before} an anomaly occurs
\citep{michael2002fault}.

Visualization methods are also valuable tools for assessing data
assumptions and can potentially be formalized as analysis validation
checks. For instance, plotting a variable's distribution using a
histogram, density plot, or bee swarm plot can reveal outliers or
deviations from normality. These visualizations could be re-framed as
analysis checks, which fail when: the data does not conform to the
visual expectation. However, translating visualization results into
binary checks remains an open challenge, requiring either manual
verification or the development of automated methods to interpret
visualization outputs. An existing example of visual test is the R
package \texttt{vdiffr} \citep{vdiffr} for graphic software unit
testing. The package saves a template plot and compares it to the
current plot to determine whether the unit tests pass or fail.

Systematically generating realistic simulated data is a key component of
our approach and is deserving of further consideration. In the PM10
example in Section~\ref{sec-pm10-mortality}, the inverse-transform
method was used to preserve the correlation structure among mortality,
PM10, and temperature. However, the simulation process can become
complex when additional restrictions are imposed or when a greater range
of scenarios is desired. In such cases, techniques like the
acceptance-reject method or permutation may be used to generate the
data. The results in Section~\ref{sec-pm10-mortality} on the broader
NMMAPS dataset suggest that our simulation procedure may have been
inadequate in reflecting the range of possible configurations that the
data could take. In particular, our split-data approach, using New York
City to guide the simulation and then applying the analysis checks to
other cities, may not have been ideal. It may be worth exploring some
recent work in data thinning \citep{neufeld2024data}, data fission
\citep{leiner2023data}, or differential privacy methods
\citep{dong_gaussian_2022} to create training datasets that are more
representative of future data. The Gaussian copula used in the example
doesn't account for the tail dependence between the variables and other
copula, e.g.~t-copula or extreme value distribution copula, could also
be explored to better capture the tail dependence.

Analysis validation checks are closely related to the concept of unit
testing in software engineering. While unit tests isolate and test
specific lines of the code, analysis validation checks focus on the
assumptions underlying the analysis rather than the explicit code
itself. Moreover, while software testing is deterministic, with clear
rules for determining failure, analysis validation checks are
probabilistic. As a result, an analysis may fail several assumption
checks yet produce an expected outcome, or pass all checks but yield an
unexpected result.

Communicating the process of data analysis is a key element to providing
transparency, improving reproducibility, and building trust with a range
of audiences. The analysis validation checks described here provide a
general way to encode the assumptions that a data analyst makes about
the data and statistical tools applied to the data. These assumptions
can be studied or challenged, depending on the specific analyst's
perspective, and can serve as a roadmap for diagnosing unexpected
results.

\section{Acknowledgement}\label{acknowledgement}

The article is created using Quarto \citep{Allaire_Quarto_2022} in R
\citep{R}. The source code for reproducing the work reported in this
paper can be found at:
\url{https://github.com/huizezhang-sherry/paper-avc}.

\section*{Supplementary material}\label{supplementary-material}
\addcontentsline{toc}{section}{Supplementary material}

The supplementary materials include a full script of the examples in the
paper (\texttt{index.R}) and its output (\texttt{index.html}), the data
used in the examples in Section 5 (\texttt{data/}), the package source
(\texttt{adtoolbox\_0.1.0.tar.gz}), and a README.md file containing the
install instructions for running the scripts.

\renewcommand\refname{References}
\bibliography{references.bib}

\end{document}